\documentclass[aps,pre,reprint,groupedaddress]{revtex4-1}
\usepackage{amsmath,amssymb}
\usepackage{bm}
\usepackage{graphicx}
\usepackage{verbatim}

\begin{document}

\title{Rheology of cohesive granular particles under constant pressure}

\author{Yuta Yamaguchi}
\email[]{yuta-y@eri.u-tokyo.ac.jp}
\affiliation{Earthquake Research Institute, University of Tokyo, 1-1-1 Yayoi, Bunkyo-ku, Tokyo 113-0032, Japan}

\author{Satoshi Takada}
\affiliation{Earthquake Research Institute, University of Tokyo, 1-1-1 Yayoi, Bunkyo-ku, Tokyo 113-0032, Japan}

\author{Takahiro Hatano}
\affiliation{Earthquake Research Institute, University of Tokyo, 1-1-1 Yayoi, Bunkyo-ku, Tokyo 113-0032, Japan}

\date{\today}

\begin{abstract}
The rheology of cohesive granular materials, under a constant pressure condition, is studied using molecular dynamics simulations. 
Depending on the shear rate, pressure, and interparticle cohesiveness, the system exhibits four distinctive phases: uniform shear, oscillation, shear-banding, and clustering. 
The friction coefficient is found to increase with the inertial number, irrespective of the cohesiveness.
The friction coefficient becomes larger for strong cohesion.
This trend is explained by the anisotropies of the coordination number and angular distribution of the interparticle forces. 
In particular, we demonstrate that the second-nearest neighbors play a role in the rheology of cohesive systems.
\end{abstract}


\maketitle


\section{Introduction}

It is important to understand the rheological properties of granular materials, for application in industry and several other fields \cite{Silbert2001, GDRMidi2004, Jop2006}.
The rheology of granular particles, under constant volume, has been intensively studied, and is known to change drastically across the jamming density \cite{Hatano2007, Olsson2007}.
Below the jamming density, the viscosity is proportional to the shear rate \cite{Bagnold1954}, and the kinetic theory is a powerful tool for understanding the rheology of dilute and moderately-dense cases \cite{Brey1998, Garzo1999, Garzo2003, Brilliantov2004, Lutsko2005, Garzo2007_1, Garzo2007_2, Mitarai2007, Chialvo2013}.
On the other hand, above the jamming density, the contacts between the particles become dominant, and the shear stress has a finite value, even for a low shear-rate limit.
Additionally, both above and below the jamming density, critical scaling is reported \cite{Hatano2007, Olsson2007}.
Another important setup for investigating the rheology is the constant pressure condition, where the relationship between the inertial number (dimensionless shear rate) and stress ratio (shear stress divided by the normal stress) has been intensively studied; this relationship is a monotonically increasing function \cite{Cruz2005, Bouzid2013, Bouzid2015, DeGiuli2015, DeGiuli2016, Barker2017}.
The relationship changes, depending on the existence of tangential friction between the particles \cite{Bouzid2013}.

Under certain situations, such as the van der Waals force for fine powders \cite{Israelachvili2011, Castellanos2005}, capillary force for moderately humid particles \cite{Rowlinson1982, Herminghaus2005, Mitarai2005, Mitarai2006, Mitarai2010, Herminghaus2013}, and electromagnetic force for magnetic beads \cite{Forsyth2001}, the cohesiveness between particles is nonnegligible.
The existence of cohesive forces alters the rheology because phase transition or nucleation occurs \cite{Heist1994, Strey1994, Yasuoka1998} and several patterns appear, depending on a set of control parameters \cite{Takada2014, Saitoh2015}.
The dissipation between collisions increases, when the temperature is comparable to the magnitude of the attractive forces \cite{Muller2011, Murphy2015, Takada2016}.
Recently, certain papers have studied the rheology of cohesive granular particles, under a constant volume condition \cite{Gu2014, Chaudhuri2012, Irani2014, Irani2016, Saitoh2015, Takada2017}. 
There exists a yield stress, even below the jamming density \cite{Irani2014, Irani2016, Iordanoff2005, Aarons2006}, which does not appear for noncohesive systems. 
Irani et al. \cite{Irani2014, Irani2016} determined a negative peak below the jamming density at constant volume conditions. 
This result completely differs from that in noncohesive cases.   
On the other hand, many experiments have been performed under a constant pressure condition, which is more realistic than those under a constant volume condition \cite{Irani2014, Irani2016}.
Therefore, it is important to study the rheology, under a constant pressure condition. 

In this paper, we study the role of cohesive interparticle forces on the rheology of model granular systems, under constant pressure conditions, using molecular dynamics (MD) simulations.
The organization of this paper is as follows:
In the next section, we explain our simulation model.
Section \ref{sec:results} is the principal part of this paper, where we present the simulation results. 
We then discuss and conclude our results in Secs.\ \ref{sec:discussion} and \ref{sec:conclusion}, respectively. 
In appendices \ref{sec:Histogram} and \ref{sec:Kurtosis}, we present the mean velocity distribution and its kurtosis to validate our criterion to distinguish the phases.

 \begin{figure}[htbp]
 \centering
  \includegraphics[width=.6\linewidth]{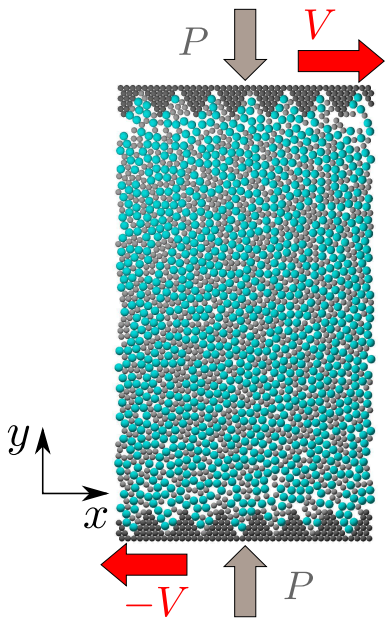}
  \caption{Simulation model.
  Bidisperse particles are confined between the top and bottom walls by pressure applied in the $y$-direction.
  We apply a shear to the system by moving the walls in the $x$-direction; the upper (lower) wall moves in the positive (negative) $x$-direction.}
  \label{fig:setup}
 \end{figure}
\section{Simulation model}
In this section, we describe our simulation model.
We consider a two-dimensional system, and prepare 2,000 bidisperse (50:50) frictionless particles with diameters and masses of $d_1(\equiv d)$, $d_2$, and $m_1(\equiv m)$, $m_2$, respectively.
In this paper, the dispersity is fixed as $d_2=1.4d$, and $m_2=1.4^2 m$.
Walls with length, $L_x=42d$,are made of smaller particles, aligned in the $x$-direction with an interval, $d$.
To efficiently apply shear, we also align the particles to form triangles with sides, $6d$, as shown in Fig.\ \ref{fig:setup}.
The upper (lower) wall is compressed with a force, $-PL_x$ ($+PL_x$), in the $y$-direction, and moves with a velocity, $V$ ($-V$), in the $x$-direction.
In our system, the system size fluctuates in the $y$-direction.
We define length, $L_y$, as the long-time average of the system size in the $y$-direction, for later discussion.

The interaction between particles is described by the sum of the elastic force from the potential and dissipative forces.
The potential part is given by
\begin{align}
	U(r_{ij})= \begin{cases}
		\epsilon \left[\left(1-\frac{r_{ij}}{d_{ij}}\right)^2 - 2u^2\right] & \frac{r_{ij}}{d_{ij}} \le 1+u, \\
		-\epsilon \left(1+2u - \frac{r_{ij}}{d_{ij}}\right)^2 & 1+u < \frac{r_{ij}}{d_{ij}} \le 1+2u,\\
		0 & \frac{r_{ij}}{d_{ij}}>1+2u,
	\end{cases}\label{eq:potential}
\end{align}
where $d_{ij}\equiv (d_i+d_j)/2$, $r_{ij}$ is the distance between the $i$-th and $j$-th particles, $\epsilon$ relates to the stiffness of the particles, and $u$ characterizes the well depth and width. 
It is to be noted that this potential (\ref{eq:potential}) is used not only for cohesive grains but also for attractive emulsions \cite{Irani2014, Lois2008, Chaudhuri2012}.
In addition, we consider the dissipative force, which acts when two particles overlap each other.
The magnitude of the dissipative force also depends on the relative velocity of the two particles.
Its explicit expression is given by
\begin{equation}
	\bm{F}^{\rm diss}(\bm{r}_{ij}, \bm{v}_{ij}) = -\zeta \Theta(d_{ij}-r_{ij}) (\bm{v}_{ij} \cdot \hat{\bm{r}}_{ij})\hat{\bm{r}}_{ij},
	\label{eq:dissipation}
\end{equation}
where $\zeta$ is the dissipation rate, $\bm{v}_{ij}\equiv \bm{v}_i - \bm{v}_j$ is the relative velocity between the particles, $\hat{\bm{r}}_{ij}$ is a unit vector parallel to $\bm{r}_{ij}=\bm{r}_i - \bm{r}_j$, and $\Theta(x)$ is a step function, where $\Theta(x)=1$ for $x>0$ and $0$ for $x\le 0$.
Then, the force acting on the $i$-th particle is expressed as
\begin{equation}
	\bm{F}_i = \sum_{j\neq i} \left[ -\bm\nabla_i U(r_{ij}) + \bm{F}^{\rm diss} (\bm{r}_{ij}, \bm{v}_{ij})\right]. 
\end{equation}
In addition, the equation of motion of the walls is selected to be overdamped with a dissipation rate, $\zeta_{\rm wall}=10 \sqrt{m\epsilon}/d$, to accelerate the simulations. 
The time evolution of the walls is described by
\begin{align}
	-PL_x - \zeta_{\rm wall}\dot y_{\rm upper} &= 0, \\
	PL_x - \zeta_{\rm wall}\dot y_{\rm lower} &= 0.
\end{align}
Here, ${y}_{\rm upper}$ (${y}_{\rm lower}$) corresponds to the coordination of the upper (lower) wall in the $y$-direction. 

In this paper, all the quantities are nondimensionalized in terms of $m$, $d$, and $\epsilon$.
As in previous studies \cite{Irani2014, Irani2016}, we select the dimensionless dissipation rate as $\zeta^* \equiv \zeta d/\sqrt{m\epsilon}=2$. 
In particular, this dissipation rate approximately corresponds to a restitution coefficient of approximately $0.135$. 
Certain metal material, such as copper or aluminum, have this restitution- coefficient value. 
It is to be noted that the choice of the restitution coefficient does not drastically affect the rheology \cite{Cruz2005}.

\section{Results}\label{sec:results}
In this section, we present the results, when three dimensionless parameters are controlled:\ the velocity of the moving walls, $V^*(\equiv V \sqrt{m/\epsilon})$, that determines the shear rate, the normal stress, $P^*(\equiv P d^2/\epsilon)$, and the strength of the attractive potential, $u$ in Eq.\ (\ref{eq:potential}). 
\subsection{Phase diagram}\label{sec:phase_diagram}
A set of control parameters specifies the behavior of the system in the steady states. 
We determined four steady states: the (i) uniform shear phase, (ii) oscillation phase, (iii) clustering phase, and (iv) shear-banding phase, as described below. 

Initially, the long-time averaged velocity profiles in the $y$-direction are linear in phases (i) and (ii). 
All the particles flow uniformly and uniform shear is applied in phase (i). 
Although the long-time averaged behavior of phase (ii) is similar to that of phase (i), the temporal behavior is completely different.
Phase (ii) has large velocity fluctuations in the bulk and backward movements are observed within a short time period.

 \begin{figure}[htbp]
  \centering
  \includegraphics[width=1.0\linewidth]{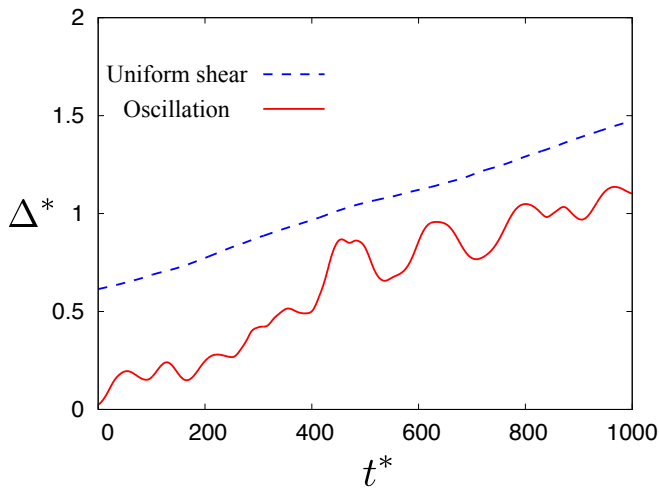}
  \caption{Time evolution of the integrated mean displacement, $\Delta$ (Eq.\ (\ref{eq:delta})), in the region near $y=L_y/4$ for phase (i) (dashed line) with $u = 2\times 10^{-4}$, $P^* = 10^{-3}$, and $V^* = 10^{-3}$, and phase (ii) (solid line) with $u = 2\times 10^{-1}$, $P^* = 10^{-3}$, and $V^* = 10^{-3}$, where $t^*\equiv t\sqrt{\epsilon/md^2}$ and $\Delta^*\equiv\Delta/d$. }
  \label{fig:slip}
 \end{figure}
To characterize phase (ii), we focus on the region ($y=L_{y}/4$) between the center of the system and the upper wall with width, $L_y/31$, and calculate the average velocity, $\bar {v}^*_x(y=L_y/4)$, in this region. 
We select this region because the long-time averaged velocity is not zero and the effect of the walls is relatively small, in this regime. 
We define the integrated mean displacement, $\Delta(t)$, in this region as
\begin{equation}
	\Delta(t)\equiv \int_0^t dt^\prime \bar{v}_x(t^\prime).
	\label{eq:delta}
\end{equation}
The time evolution of $\Delta(t)$ in phases (i) and (ii) is shown in Fig.\ \ref{fig:slip}.
The oscillation phase (ii) has large fluctuations and the behavior of this phase is clearly different from that of phase (i). 
The intermittent decrease of $\Delta(t)$ indicates a backward motion for reducing the deformation energy by the shear. 

 \begin{figure}[htbp]
  \centering
  \includegraphics[width=1.0\linewidth]{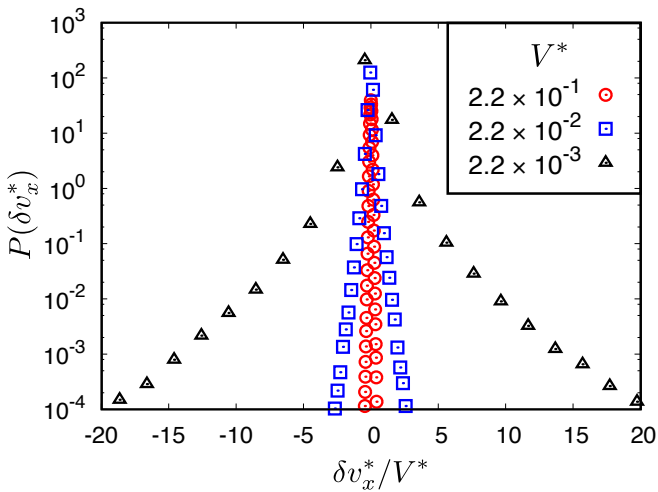}
  \caption{Distribution function of the velocity fluctuation $\delta v_x^* \equiv v_x -  \bar{v}_x$ for various velocity $V$ with $v_x^*\equiv v_x \sqrt{m/\epsilon}$. 
  Here, an average flow $\bar{v}_x$ is subtracted from the particle velocity $v_x$. 
  The horizontal axis is the normalized particle velocity divided by the velocity of moving walls, $V^*$. 
  We use the following parameters ($u = 2\times 10^{-2}, P^* = 10^{-2}$, and $V^* = 2.2\times10^{-1}$ (circles), $2.2\times 10^{-2}$ (squares), $2.2\times10^{-3}$ (triangles)). 
  We classify $V^* = 2.2\times10^{-1}, 2.2\times 10^{-2}$ as phase (i), and $2.2\times10^{-3}$ as phase (ii). }
  \label{fig:particle_velocity}
 \end{figure}
To specify the difference between phases (i) and (ii) qualitatively, we focus on the distribution function of the velocity fluctuation of the particles. 
Here, we measure this distribution in the region between $y=L_y/4$ and $y=-L_y/4$ to reduce the effect of walls, and subtract the average velocity $\bar{v}_x(y)$ from the particle velocity. 
Figure \ref{fig:particle_velocity} exhibits the exponential distributions for wide range of the wall velocities $V$, which dependence is reported for cohesive systems in the previous study \cite{Takada2014}.
Here, the velocity fluctuation becomes larger than the wall velocity as the wall velocity decreases (for instance, $V^*=2.2\times 10^{-3}$ in Fig.\ \ref{fig:particle_velocity}).
This large fluctuation also suggests the existence of the backward motion (phase (ii)) as shown in Fig.\ \ref{fig:slip}.
To classify the phase (ii) from the phase (i), we use the following threshold.
If the probability for $|\delta v_x^*/V^*|>1$ is more than $0.018$, the phase belongs to the oscillation phase (ii).
This value corresponds to the probability that the value deviates from three times larger than the standard deviation of the exponential distribution.
The classification of the phases (i) and (ii) is shown later in this subsection.
It is noted that our choice of the criterion is reasonable because we have obtained the similar phase diagrams when we use different criterion based on the mean velocity analysis (see Appendix \ref{sec:Histogram}).

 \begin{figure}[htbp]
  \includegraphics[width=1.0\linewidth]{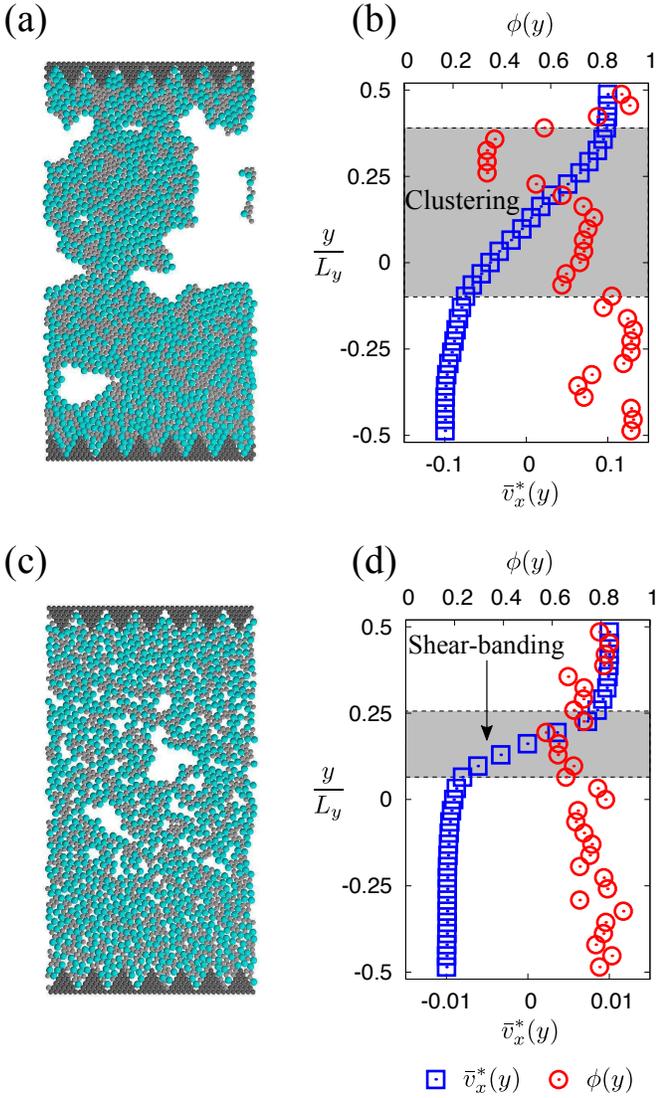}
  \caption{(a) Snapshot and (b) density (circles) and velocity (squares) profiles of the clustering phase ($u = 2\times 10^{-1}, P^* = 10^{-3}$, and $V^* = 10^{-1}$). 
  Similar plots are shown in (c) and (d) for the shear-banding phase, where we use the following parameters ($u = 2\times 10^{-2}, P^* = 10^{-3}$, and $V^* = 10^{-2}$).
  The shaded regions in (b) and (d) exhibit clustering and shear-banding, respectively.}
  \label{fig:Profile}
 \end{figure}

Next, we describe the other phases. 
When the normal stress is low, and the attractive potential becomes dominant with respect to the shear, uniform shear cannot be applied to the system, even in the long-time average. 
In this case, there exist two characteristic phases: the (iii) clustering and (iv) shear-banding phases. 
Typical snapshots, and the density and velocity profiles of these phases are depicted in Fig.\ \ref{fig:Profile}. 
In phase (iii), certain clusters are formed and they roll over with time, in the bulk region (see Fig.\ \ref{fig:Profile}(a)).
Uniform shear cannot be applied due to the existence of clusters, as shown in Fig.\ \ref{fig:Profile}(b).
It is to be noted that the packing fraction also decreases in this region because voids exist near the clusters.
On the other hand, in phase (iv), the packing fraction remains nearly uniform, as shown in Fig.\ \ref{fig:Profile}(d), while the shear is localized in a certain region (see Fig.\ \ref{fig:Profile}(d)). 
Note that this localization occurs not only in the bulk region but also in the region near the walls, depending on the parameters. 

 \begin{figure}[htbp]
 \begin{center}
  \includegraphics[width=1.0\linewidth]{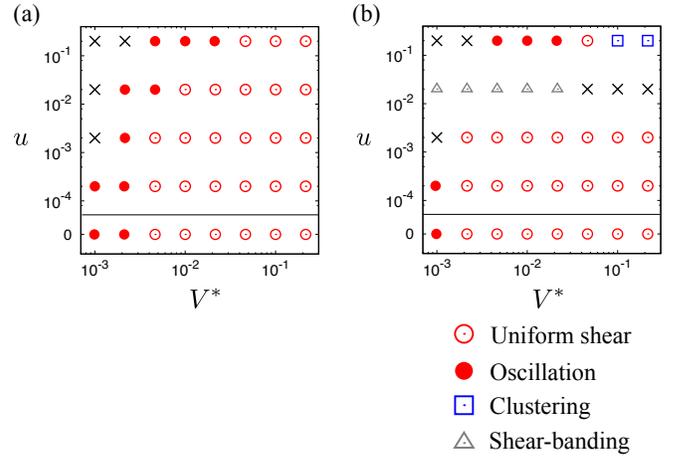}
  \caption{Phase diagrams for (a) $P^*=10^{-2}$, (b) $10^{-3}$. 
  There are four phases: (i) uniform shear (open circles), (ii) oscillation (filled circles), (iii) clustering (open squares), and (iv) shear-banding (open triangles), where the dimensionless normal stress is $P^*\equiv Pd^2 /\epsilon$ and the dimensionless velocity of the moving walls is $V^*\equiv V\sqrt{m/\epsilon}$. 
  We plot cross marks for the others.}
  \label{fig:phase}
  \end{center}
 \end{figure}

Based on the above discussion, we present the phase diagrams for the various normal stresses, $P$, in Fig.\ \ref{fig:phase}.
The uniform shear phase is stable, when the velocity of the moving walls is more. 
On the other hand, the oscillation phase appears, when the velocity is less. 
When the attractive potential is strong, the oscillation phase appears, even in the faster regime, indicating that the attractive force controls the oscillation phase. 
Only in the region where the attraction is dominant compared to the repulsion, the clustering and shear-banding phases emerge.

\begin{figure}[htbp]
  \includegraphics[width=1.0\linewidth]{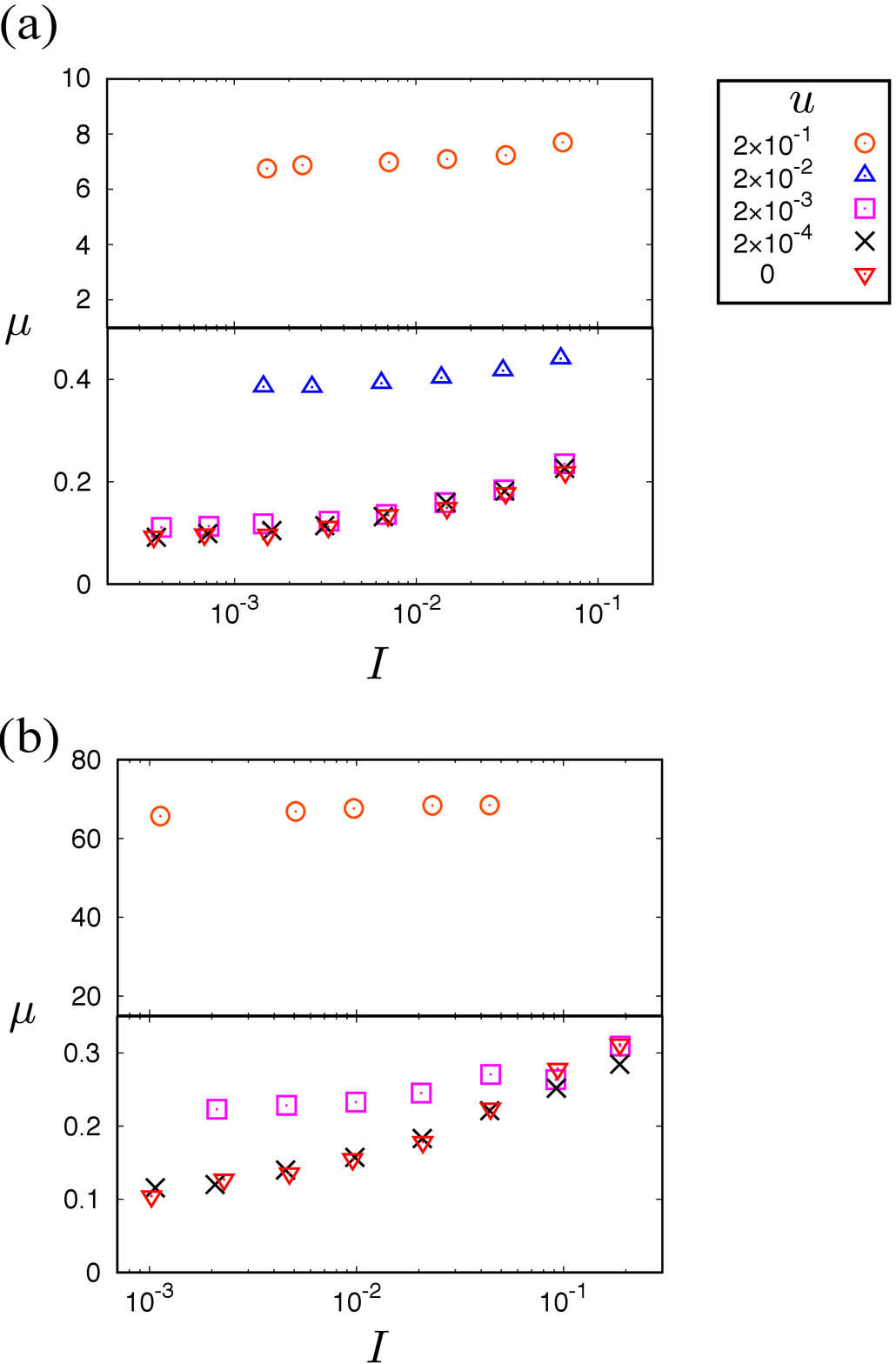}
  \caption{Plot of the $\mu-I$ rheology for various $u$: $u=2\times10^{-1}$ (circles), $2\times10^{-2}$ (triangles), $2\times10^{-3}$ (squares), $2\times10^{-4}$ (cross marks), and $0$ (reverse triangles), when the normal stress is (a) higher ($P^*=10^{-2}$) and (b) lower ($P^*=10^{-3}$).}
  \label{fig:mu_I_rheology_u}
\end{figure}
\subsection{Flow curve}\label{sec:Flow curve}
Next, we present the $\mu-I$ curves for the system. 
Here, $\mu(\equiv -\sigma_{xy}/P)$ is the friction coefficient, and $I$ is the inertial number defined by $I\equiv\dot{\gamma}\sqrt{m/(Pd)}$. 
We define the shear rate, $\dot{\gamma}$, as the slope of the velocity profile in the region, where the velocity profile is linear in the long-time average. 
Hence, we focus only on the (i) uniform shear and (ii) oscillation phases. 
The shear stress, $\sigma_{xy}$, is the $xy$ component of the microscopic pressure tensor defined by
\begin{equation}
	\sigma_{\alpha \beta} = \frac{1}{L_x L_y}\sum_{i=1}^{N} \left(m_i V_{i,\alpha}V_{i,\beta} + \frac{1}{2}\sum_{j\neq i} r_{ij,\alpha} F_{ij,\beta}\right),
	\label{eq:shear stress}
\end{equation}
where $\bm{V}_{i} \equiv \bm{v}_{i}-\bm{U}(y)$ is the deviation from the macroscopic velocity field, $\bm{U}(y)$, each time.

Figure \ref{fig:mu_I_rheology_u} shows the $\mu-I$ rheology for cases with (a) higher normal stress ($P^*=10^{-2}$) and (b) lower normal stress ($P^*=10^{-3}$), when the strength of the attraction, $u$, is varied. 
Initially, there exists a yield stress in the low shear limit. 

Next, the curves collapse in the high shear-rate regime, which is independent of the attractive potential, $u$, except for $u=2\times 10^{-1}$.
This collapse indicates that the effect of attraction is negligible. 
On the other hand, in the low shear-rate regime, the effect of attraction is considerable, i.e., the friction coefficient increases as the attraction becomes high. 
In contrast, the flow curve for the strong attractive potential ($u=2\times 10^{-1}$) is completely different from that of the weak attraction cases.
The friction coefficient is abnormally large, as shown in Figs. 6(a) and (b), and this trend is notable for the low normal-stress cases, where the attraction is dominant. 
This is because the system can be supported by attraction even in the case without compressive force, and the low normal stress gives the high friction coefficient. 
This large friction coefficient is also explained from different points of view in a later subsection.

\begin{figure}[htbp]
  \includegraphics[width=1.0\linewidth]{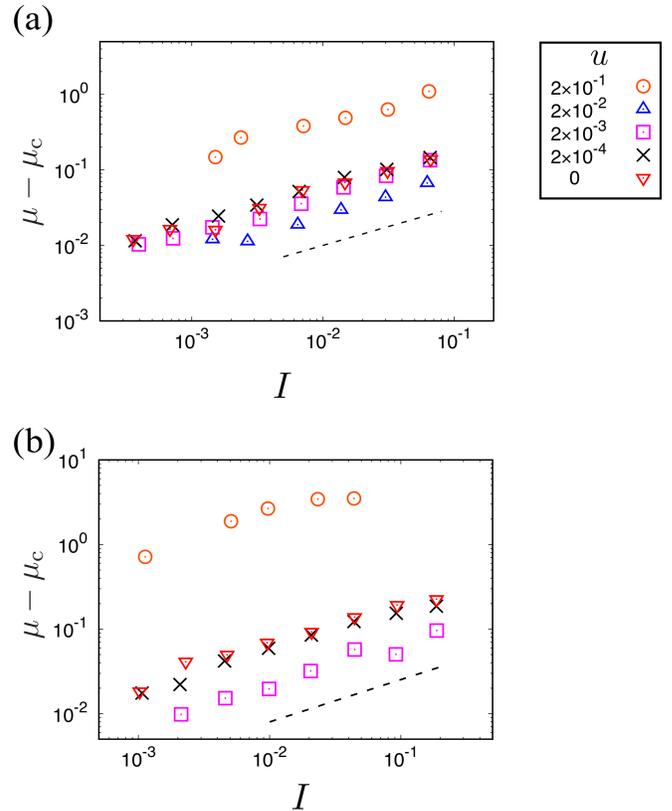}
  \caption{Plot of the $\mu-\mu_\mathrm{c}$ curve for various $u$, when the normal stress is (a) higher ($P^*=10^{-2}$) and (b) lower ($P^*=10^{-3}$). 
  The symbols are the same as in Fig.\ \ref{fig:mu_I_rheology_u}.
  $\mu_{\rm c}$ are the plateau values in the low shear-rate limit, in Fig.\ \ref{fig:mu_I_rheology_u}. 
  The dashed lines represent Eq. (\ref{eq:mu-muc_I}).}
  \label{fig:mu-mu_c_u}
\end{figure}
To investigate the effect of cohesive forces on the flow curve, we first focus on the dynamic part of the friction coefficient, i.\ e., we subtract the plateau value, $\mu_{\rm c}$, from the flow curve shown in Fig.\ \ref{fig:mu_I_rheology_u}, for the case where the attraction is dominant. 
Figure \ref{fig:mu-mu_c_u} shows the plot of $\mu-\mu_\mathrm{c}$ against the inertial number, $I$, for various $P$ and $u$.
For all the parameters, these data can be fitted using 
\begin{equation}
	\mu = \mu_\mathrm{c}+a \sqrt{I}, \label{eq:mu-muc_I}
\end{equation}
with the fitting parameters, $a$ and $\mu_{\rm c}$. 
It should be noted that this result coincides with those of the earlier studies \cite{Cruz2005, Bouzid2013, Bouzid2015, DeGiuli2015, DeGiuli2016, Barker2017}, which reported this relationship using frictionless particles without cohesive interactions, under a constant pressure condition. 
This implies that the cohesiveness does not affect the exponent of $I$ in Eq.\ (\ref{eq:mu-muc_I}), and affects only the static friction, $\mu_{\rm c}$, and the coefficient, $a$.

\begin{figure}[htbp]
  \includegraphics[width= 0.85\linewidth]{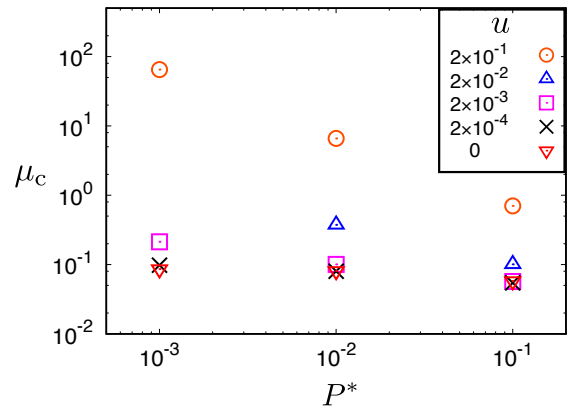}
  \caption{Plateau value, $\mu_\mathrm{c}$, in the flow curve (Fig.\ \ref{fig:mu_I_rheology_u}) as a function of the normal pressure, $P^*$, for various $u$. 
  The symbols are the same as in Fig.\ \ref{fig:mu_I_rheology_u}}
  \label{fig:mu_c-P}
\end{figure}
Furthermore, we demonstrate that the static part of the friction coefficient, $\mu_\mathrm{c}$, for various normal stresses and cohesiveness, in Fig.\ \ref{fig:mu_c-P}, exhibits that $\mu_{\rm c}$ is a decreasing function of the normal stress.
In addition, it is to be noted that $\mu_{\rm c}$ tends to be independent of the pressure, when the cohesion becomes weak.
Then, we can conclude that the cohesive force affects the relationship between the static friction coefficient and the normal pressure. 
\begin{figure}[htbp]
  \includegraphics[width=0.9\linewidth]{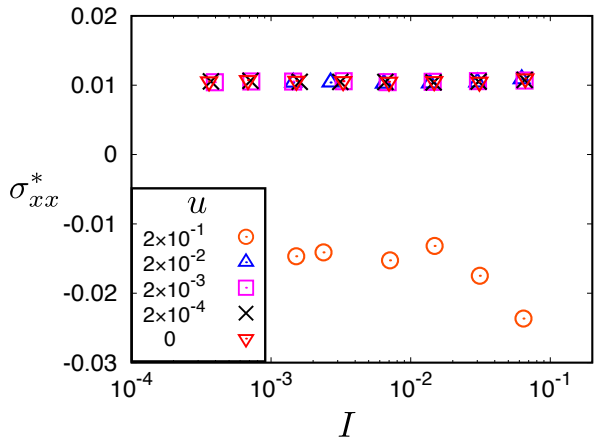}
  \caption{Normal stress, $\sigma_{xx}^*(\equiv \sigma_{xx}d^2/\epsilon)$, in the $x$-direction versus the inertial number for various $u$, when $P^*=10^{-2}$.
  The symbols are the same as in Fig.\ \ref{fig:mu_I_rheology_u}.}
  \label{fig:sigmaxx}
\end{figure}
We also plot the normal stress, $\sigma_{xx}$, in the $x$-direction as a function of the inertial number, as shown in Fig.\ \ref{fig:sigmaxx}. 
For the weak attraction case, the normal stress in the $x$-direction is equivalent to the normal stress, $P^*=10^{-2}$. 
In contrast, it becomes lower than the normal stress and tends to be negative for the attraction dominant case, $u=2\times 10^{-1}$. 
It is to be noted that the normal stress, $\sigma_{yy}$, in the $y$-direction is always positive, even in this regime. 
This suggests the existence of an anisotropy in the system, when the attraction is dominant, which is discussed in a later subsection.

\subsection{Packing fraction}\label{sec:Packing fraction}
 \begin{figure}
  \centering
  \includegraphics[width=1.0\linewidth]{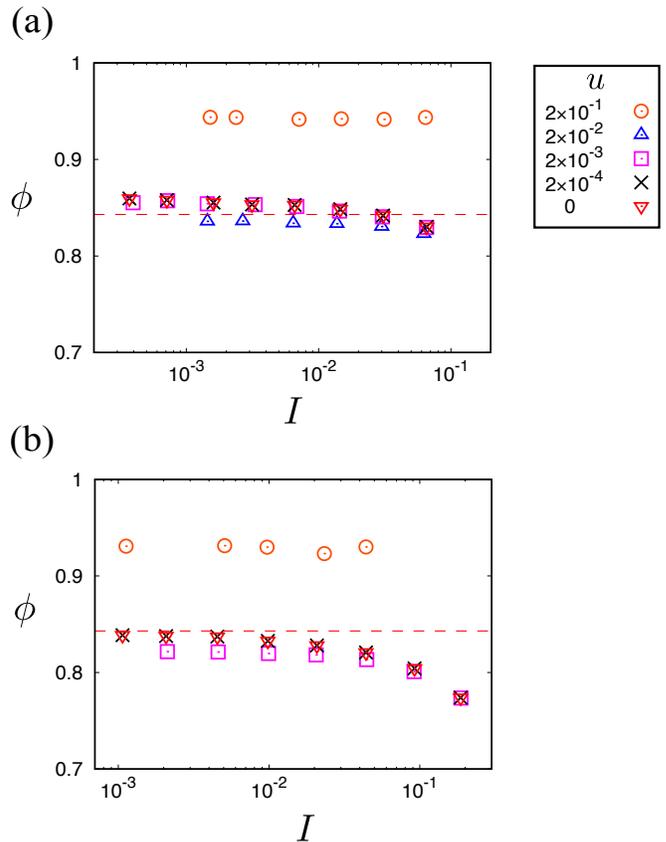}
  \caption{Packing fraction for various $u$, when the normal stress is (a) higher ($P^*=10^{-2}$) and (b) lower ($P^*=10^{-3}$). 
  The symbols are the same as in Fig.\ \ref{fig:mu_I_rheology_u}.
  The dashed lines indicate the jamming density ($\phi_{\rm J}=0.843$) for a purely repulsive system.}
  \label{fig:packing fraction}
 \end{figure}
Under a constant pressure condition, we cannot control the packing fraction, which is determined by a set of control parameters. 
We plot the inertial number dependence of the packing fraction in Fig.\ \ref{fig:packing fraction}. 
The packing fraction is nearly independent of the inertial number but depends on the cohesiveness and the pressure.
When the cohesiveness is strong, the packing fraction is nearly equal to the jamming density, as shown by the dashed line in Fig.\ \ref{fig:packing fraction}. 
On the other hand, a high-fraction is realized for $u=2\times10^{-1}$.
Note that high pressure also makes the system denser. 
The parameters for this high packing fraction are same as those for the abnormally large friction coefficient discussed in the previous subsection.  
In particular, a large friction coefficient is achieved in the low pressure and strong attraction case. 

\subsection{Anisotropy}\label{sec:Anisotropy}
\begin{figure}[htbp]
  \includegraphics[width=0.85\linewidth]{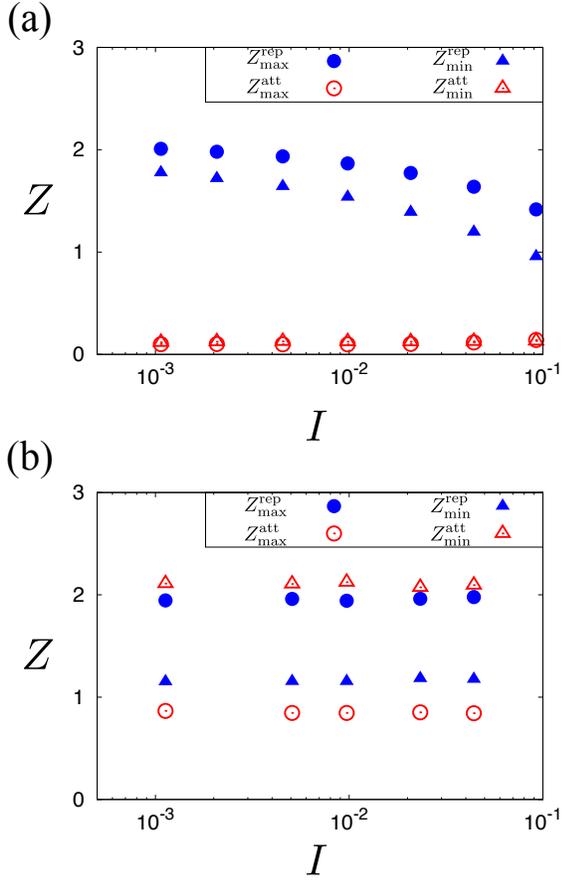}
  \caption{Anisotropic coordination number versus the inertial number for the (a) repulsion dominant ($u=2\times 10^{-4}$ and $P^*=10^{-3}$) and (b) attraction dominant ($u=2\times 10^{-1}$ and $P^*=10^{-3}$) cases. 
  The anisotropies of the coordination number are represented by the direction: the maximum compression part ($Z_{\rm max}$) or the minimum compression part ($Z_{\rm min}$). 
  The filled and open circles represent the coordination numbers of the maximum compression part in the repulsive and attractive ranges, respectively.
  The filled and open triangles indicate the coordination number of the minimum compression part in the repulsive and attractive ranges, respectively.}
  \label{fig:coordination number}
\end{figure}
To clarify the reason for the anisotropies of $\sigma_{xx}$ and $\sigma_{yy}$, we focus on the microscopic properties, in particular, the coordination number and interparticle forces. 
First, we present the anisotropic coordination numbers in Fig.\ \ref{fig:coordination number}. 
Here, we define the coordination number, $Z$, as the mean number of particles, which interact with a certain particle. 
In addition, we divide the coordination number into repulsive ($Z^{\rm rep}$) and an attractive ($Z^{\rm att}$) parts, from Eq.\ (\ref{eq:potential}). 
Moreover, in our sheared system, there are two principal axes \cite{Barker2017}: the maximum compressional axis, which corresponds to the direction of $\theta=-\pi/4$, and the minimum compressional axis, which corresponds to the direction of $\theta=\pi/4$.
Here, $\theta$ is defined as the angle from the $x$-axis, counterclockwise.
Therefore, we divide $Z^{\rm rep}$ ($Z^{\rm att}$) into the maximum compressions part, $Z^{\rm rep}_{\rm max}$ ($Z^{\rm att}_{\rm max}$) (second and fourth orthants, i.e., $\pi/2\le \theta<\pi$ and $3\pi/2\le \theta<2\pi$), and the minimum compression part, $Z^{\rm rep}_{\rm min}$ ($Z^{\rm att}_{\rm min}$) (first and third orthants, i.e., $0\le \theta<\pi/2$ and $\pi\le \theta<3\pi/2$). 
The weak cohesion case is shown in Fig.\ \ref{fig:coordination number}(a) ($u = 2\times 10^{-4}, P^* = 10^{-3}$). 
Although $Z_{\rm max}^{\rm rep}$ is slightly larger than $Z_{\rm min}^{\rm rep}$, there are hardly any differences in the anisotropic properties, and the coordination number in the attraction range is negligible. 
On the other hand, Fig.\ \ref{fig:coordination number}(b) shows the case, where the attraction is dominant. 
Here, we can observe anisotropies, such as $Z_{\rm max}^{\rm rep}>Z_{\rm min}^{\rm rep}$ and $Z_{\rm min}^{\rm att}>Z_{\rm max}^{\rm att}$. 
These results are as predicted because a high value of $Z_{\rm max}^{\rm rep}$ represents repulsion-dominance in the maximum compressional axis, and a high value of $Z_{\rm min}^{\rm att}$ represents attraction-dominance in the minimum compressional axis. 

\begin{figure}
  \includegraphics[width=0.8\linewidth]{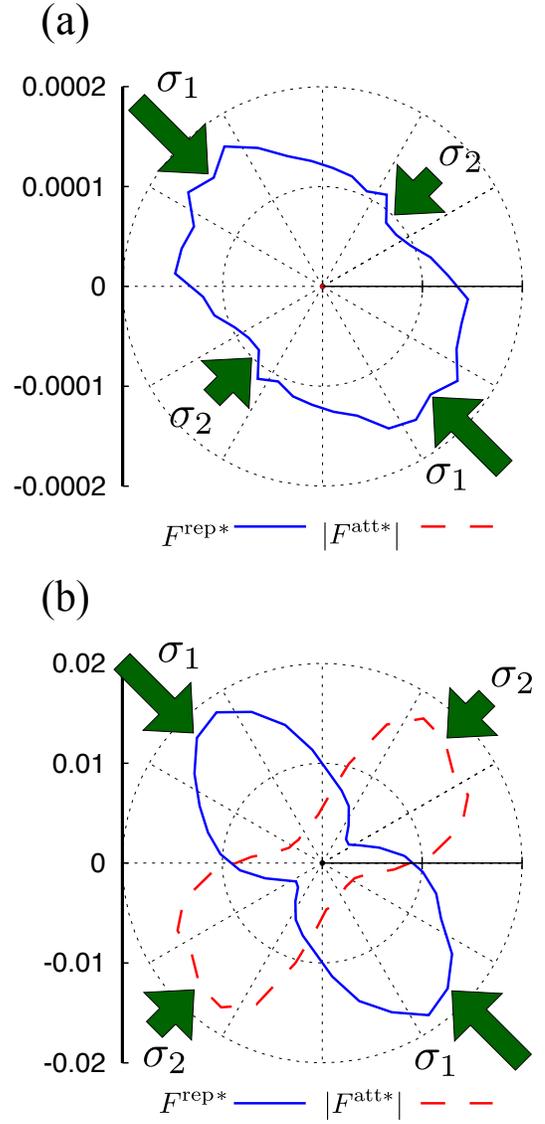}
  \caption{Angular distribution of the interparticle interactions for the (a) weak cohesion ($u = 2\times 10^{-4}, P^* = 10^{-3}, V^* = 2.2\times10^{-3}$) and (b) strong cohesion ($u = 2\times 10^{-1}, P^* = 10^{-3}, V^* = 2.2\times10^{-2}$) cases. 
   The solid and dashed lines represent the repulsive and attractive interactions, respectively.
   Here, the absolute values are plotted, for the attractive forces. 
   The arrows indicate the maximum and minimum compression axes ($\sigma_1$ and $\sigma_2$), respectively.
   For case (a), it is to be noted that the attractive force is invisible because its contribution is considerably weaker than the repulsive force.}
  \label{fig:aniso_force}
\end{figure}

Additionally, we investigate the angular distribution of the interparticle forces to reveal further details on the anisotropy. 
These distributions for the weak-cohesion and cohesion-dominant cases are shown in Figs.\ \ref{fig:aniso_force}(a) and (b), respectively. 
Here, we present the absolute values of the repulsive and attractive forces for visibility. 
As previously mentioned, the maximum compressional axis ($\theta = -\pi / 4$) coincides with the repulsion-dominant region. 
On the other hand, for the attraction dominant region, the attractive force becomes maximum at the minimum compressional axis ($\theta = \pi / 4$), as shown in Fig.\ \ref{fig:aniso_force}(b). 

\section{Discussion}\label{sec:discussion}
\begin{figure}
  \includegraphics[width=0.8\linewidth]{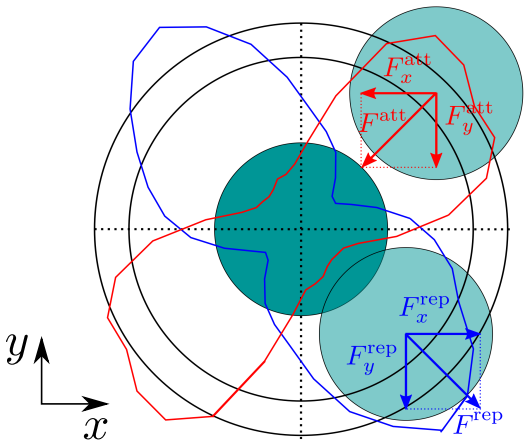}
  \caption{Schematic of the decomposition of the repulsive and attractive forces in Fig.\ \ref{fig:aniso_force}(b). 
  }
  \label{fig:aniso_force_diss}
\end{figure}
First, we discuss the abnormally high friction coefficient, when the attractive potential is strong and the normal stress is low, as shown in Fig.\ \ref{fig:mu_I_rheology_u}(b).
It is to be noted that the normal stress in the $x$-direction, $\sigma_{xx}$, also becomes large in this regime, simultaneously.
We interpret these phenomena schematically. 
Figure \ref{fig:aniso_force_diss} shows the forces acting on a fixed particle.
We decompose the repulsive (attractive) force vector, $\bm F^{\rm rep}$ ($\bm F^{\rm att}$), into $\bm F_x^{\rm rep}$, and $\bm F_y^{\rm rep}$ ($\bm F_x^{\rm att}$ and $\bm F_y^{\rm att}$) in the case of Fig.\ \ref{fig:aniso_force}(b) (see Fig.\ \ref{fig:aniso_force_diss}). 
Here, $\bm F_x^{\rm rep}$ and $\bm F_x^{\rm att}$ have opposite signs, causing a decrease in $\sigma_{xx}$, compared to the purely repulsive system, as shown in Fig.\ \ref{fig:sigmaxx}.
On the other hand, $\bm F_y^{\rm rep}$ and $\bm F_y^{\rm att}$ have the same sign and the sum of both increases $\sigma_{xy}$, as per the second term in Eq.\ (\ref{eq:shear stress}).
This is the origin of the abnormally large friction coefficient.

Next, we discuss the high packing fraction presented in Sec.\ \ref{sec:Packing fraction}.
When the normal stress is high, and the cohesive force is strong, the packing fraction is approximately $1.6$, which is nearly twice larger than the jamming density for purely repulsive systems.
To clarify the origin of this high packing fraction, we consider the energy balance.
We focus on a particle in the bulk region. 
Here, we only consider the region between this particle and a wall (we select the upper wall, for simplicity) because of the axial symmetry with respect to the $x$-axis. 
Between this particle and the wall, there exist approximately two and three particles as the first and second nearest neighbors, respectively, as per the contact number discussed in Sec.\ \ref{sec:Anisotropy}.
From the energy balance, the following relationship is approximately satisfied:
\begin{equation}
	Pd \cdot \delta \sim \left( 2\cdot \frac{\epsilon}{d^2}\delta^2 - 3\cdot 2\epsilon u^2 \right ) -\left(- 2\cdot 2\epsilon u^2 \right ), \label{eq:energy_balance}
\end{equation}
where $\delta$ is the mean overlap and $2\epsilon/d^2$ corresponds to the linear spring constant of the interaction (see Eq.\ (\ref{eq:potential})).
The left-side of Eq.\ (\ref{eq:energy_balance}) corresponds to the work from the normal stress; 
whereas, the right-side of Eq.\ (\ref{eq:energy_balance}) is the energy balance, before and after the deformation of the particles.
Here, the terms in the first bracket are the deformation energy from the first nearest neighbors and the attractive potential energy from the second nearest neighbors after deformation, respectively, and the second term is the attractive potential energy from the first nearest neighbors, before deformation. 
From Eq.\ (\ref{eq:energy_balance}), the overlap, $\delta$, can be estimated as
\begin{equation}
	\delta \sim \frac{1}{2}\left[\frac{Pd^3}{2\epsilon}+\sqrt{\left(\frac{Pd^3}{2\epsilon}\right)^2+4 u^2 d^2}\right],
\end{equation}
as a function of the normal stress, $P$ and the cohesiveness, $u$.
Using this overlap length, the deformation area of the particle, after deformation, is approximately given by $\Delta S\sim \delta^2 / 2$.
The packing fraction should be the jamming density for the deformed particles, and the simplest estimation is given by 
\begin{equation}
	\phi_{\rm J}^{\rm eff} = \phi_{\rm J}\times \frac{S}{S-4\Delta S}.\label{eq:phi_eff}
\end{equation}
Here, $S\equiv \pi d^2/4$ is the area of the particle, before deformation, and $S/(S-4\Delta S)$ is the deformation ratio of the particle, where factor $4$ in the denominator on the right-side is due to the existence of four particles as nearest neighbors, including other particles on the other side.
For $P^*=10^{-3}$ and $u=2\times 10^{-1}$, this density becomes $\phi_{\rm J}^{\rm eff}\sim 0.94$.
Similarly, for $P^*=10^{-3}$ and $u=2\times 10^{-4}$, the density is estimated as $\phi_{\rm J}^{\rm eff}\sim 0.84$.
These estimated values are consistent with those of the simulations observed in Fig.\ \ref{fig:packing fraction}; however, our estimation indicates that the finite deformations are nonnegligible and the packing fraction is affected by both the pressure and cohesiveness.
In particular, the existence of the second nearest neighbors play a crucial role in the large packing friction.
The system is more stable, when the second nearest neighbors enter the potential well, causing a high packing fraction, as shown in Fig.\ \ref{fig:packing fraction}.
This discrepancy may be due to the geometry of the particles, i.\ e., the contact number depends on the set of parameters, as shown in Fig.\ \ref{fig:coordination number}. 
Quantitative treatment considering the geometry can be possible, but this will be done elsewhere.

Third, we determined clustering and shear-banding phases, when the normal stress is low, as shown in Fig.\ \ref{fig:Profile}.
In phases (iii) and (iv), the shear is localized in a certain region.
In phase (iii), the existence of a cluster due to strong cohesion reduces the energy, while the rolling of this cluster increases the energy.
On the other hand, in phase (iv), shear-banding is selected instead of clustering.
These phases may be determined to minimize the energy of the system.
The parameter dependence of the phase selection is significant, but this is also intended in future. 

Finally, we compare our results with previous works \cite{Irani2014, Irani2016}. 
We prepared finite walls and moved these walls to apply a shear to the system, under a constant pressure condition. 
However, previous studies involved a constant volume condition with the Lees-Edwards boundary condition \cite{Lees1972}. 
They demonstrated that the flow curve was nonmonotonic below the jamming density, whereas it monotonically increased with the shear rate, above the jamming density.
Under a constant pressure condition, we present a monotonically increasing flow curve, irrespective of the density.
This difference may be due to the stability of the voids or droplets.
In previous studies \cite{Irani2014, Irani2016}, these droplets may survive under a certain condition below the jamming density, which is also reported in Ref.\ \cite{Takada2014}.
In contrast, droplets tend to vanish in our system because the normal stress in the $y$-direction may generally inhibit the spatial heterogeneity of the density.
In future, we intend to examine the stabilities of the droplets more carefully. 

\section{Conclusion}\label{sec:conclusion}
We performed MD simulations of cohesive granular particles, under a constant pressure condition, and sheared the system by moving the walls. 
We considered the effect of cohesion on rheology by adopting the attractive potential. 

First, we established four distinct phases, depending on the constant pressure, shear velocity of the moving walls, and the attractive potential. 
In the region, where uniform shear was applied in the long-time average, we classified the uniform shear phase and oscillation phase by investigating the distribution function quantitatively. 
In addition, when the cohesive force was strong, it was difficult to obtain uniform shear in the bulk, even if a long-time average was available. 
In such cases, we found the clustering and shear-banding phases. 
By parameter studies, we generated the phase diagram. 

Next, based on this phase diagram, we plotted the flow curve ($\mu-I$ rheology) in the region, where uniform shear was applied in the long-time average. 
The flow curves are a monotonically increasing function of the inertial number. 
By subtracting the plateau value, $\mu_{\rm c}$, from the $\mu-I$ rheology, we analyzed the exponent of $\mu-\mu_{\rm c}$ to $I$. 
This exponent, which coincides with $1/2$ when the curve was properly fit, 
is known in noncohesive systems \cite{Bouzid2013}. 
Then, we conclude that the effect of cohesion can be represented by $\mu_{\rm c}$.  

In the flow curves, strong cohesion yields a large friction coefficient due to the attractive force. 
In this region, there appear anisotropies of the coordination number and angular distribution of the interparticle forces.
These demonstrate that the repulsive forces are maximum in the maximum compressional axis, whereas the attractive forces are maximum in the minimum compressional axis, which is the origin of the large friction coefficient.

\section*{Acknowledgements}
We thank Michio Otsuki, Ryohei Seto, Walter Kob, and Hisao Hayakawa for their useful comments.
This study was supported by the following: JSPS KAKENHI (JP16H06478) in Scientific Research on Innovative Areas ``Science of Slow Earthquakes", ``Exploratory Challenge on Post-K computer" (Frontiers of Basic Science: Challenging the Limits) by the MEXT, Earthquake and Volcano Hazards Observation and Research Program by the MEXT, and the Earthquake Research Institute cooperative research program. 
The numerical computation in this study was partially carried out by the computer systems of the Earthquake and Volcano Information Center of the Earthquake Research Institute, University of Tokyo.

\appendix
\section{Histogram of the mean velocity}\label{sec:Histogram}
 \begin{figure}[htbp]
  \centering
  \includegraphics[width=1.0\linewidth]{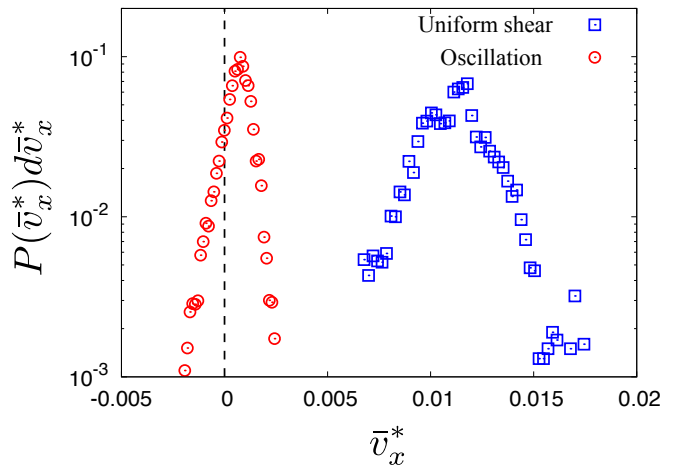}
  \caption{Distribution function of the mean velocity, $\bar{v}_x^*(\equiv\bar{v}_x \sqrt{m/\epsilon})$, of the region near $y=L_y/4$ for the (i) uniform shear phase(squares) with $u = 2\times 10^{-4}$, $P^* = 10^{-2}$, and $V^* = 2.2\times10^{-2}$, and (ii) oscillation phase(circles) with $u = 2\times 10^{-4}$, $P^* = 10^{-2}$, and $V^* = 10^{-3}$. 
  }
  \label{fig:histogram}
 \end{figure}
In this appendix, we introduce another criterion to distinguish phases (i) and (ii). 
Figure \ref{fig:histogram} shows the histograms of the average velocity $\bar{v}^*_x(y=L_y/4)$, which is discussed in Section \ref{sec:phase_diagram}. 
It is noted that this figure differs from Fig.\ \ref{fig:particle_velocity}. 
This focus on the average velocity and average flow is not subtracted from the particle velocity. 
When the mean velocity of the moving walls is sufficiently fast, the ratio of the negative velocity is negligible. 
Then, this case can be classified as phase (i). 
In contrast, when the velocity is slow, the ratio of the negative velocity becomes larger and nonnegligible. 
This indicates that at a certain instance, the mean velocity in the region ($y=L_{y}/4$) has a sign opposite to that of the upper wall; then, backward movements occur, as shown in Fig.\ \ref{fig:slip}.
To specify phase (ii) quantitatively, we can also adopt the following criterion: if the probability of the negative velocity is greater than $0.159$, the phase is (ii), which is oscillation, else the phase is (i), which is the uniform shear phase. 
This value is half the probability that the value deviates from the standard deviation of the Gaussian distribution. 
The validity of this criterion is discussed in Appendix\ref{sec:Kurtosis}, in addition. 
There is no significant difference between this criterion and the one in Section \ref{sec:phase_diagram}. 
Only a few points near the border of phases (i) and (ii) may change depending on these criteria.

\section{Kurtosis of the distribution}\label{sec:Kurtosis}
\begin{figure}
  \centering
  \includegraphics[width=1.0\linewidth]{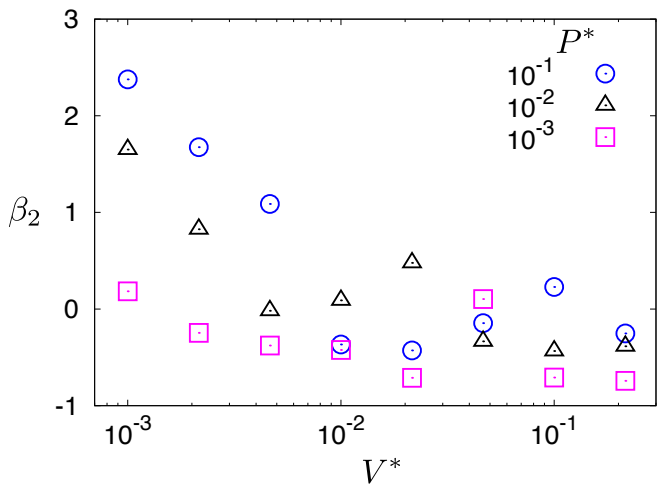}
  \caption{Kurtosis of the mean velocity in region, $y=L_y/4$, for the weak attraction, $u=2\times 10^{-4}$, when the normal stress, $P^*$, is varied: $P^*=10^{-1}$ (circles), $10^{-2}$ (triangles), and $10^{-3}$ (squares). 
  }
  \label{fig:kurtosis}
\end{figure}
In this appendix, we validate the criterion, which determines the phases.
Figure \ref{fig:kurtosis} shows the kurtosis, $\beta_2=\mu_4/\mu_2^2 - 3$, of the probability distribution versus $V^*$ in the case of weak cohesion, where $\mu_n$ is the $n$-th moment of the probability function defined by $\mu_n \equiv \langle (\bar{v}_x - \langle \bar{v}_x \rangle)^n \rangle$ with $\langle\cdots \rangle\equiv (1/N)\sum_{i=1}^N \cdots$.
It is to be noted that this criterion value, $0.159$, used in Sec.\ \ref{sec:phase_diagram} is reasonable, if the probability satisfies the Gaussian because the probability of deviating from the distribution of the standard deviation is $1-{\rm erf}(1/\sqrt{2})\simeq 0.317$, with the error function, ${\rm erf}(x)$.
When the shear is dominant, the kurtosis is approximately $0$, indicating that the probability function is well fitted by the Gaussian, while the kurtosis is approximately $3$ for the lower-shear regime, indicating that the probability function is exponential. 
Even in this regime, we use the same criterion because the probability of the value deviating from the standard deviation of the exponential distribution is $\exp(-\sqrt{2})\simeq 0.243$, which is slightly smaller than the value for the Gaussian, but is qualitatively reasonable even in this regime.





\end{document}